\documentclass[10pt]{article}

\usepackage[english]{babel}

\usepackage[dvips]{graphics,color}
\usepackage{newlfont,epsfig}

\usepackage{graphicx}

\begin{document}
\date{}

\begin{center}
{\Large\bf Quantum key distribution using bright polarized coherent states}
\end{center}
\begin{center}
{\large L.F.M. Borelli \footnote{borelli@ifi.unicamp.br} and A. Vidiella-Barranco \footnote{vidiella@ifi.unicamp.br}}
\end{center}
\begin{center}
{\bf\normalsize{ Instituto de F\'\i sica ``Gleb Wataghin'' - Universidade Estadual de Campinas}}\\
{\bf\normalsize{ 13083-970   Campinas  SP  Brazil}}\\
\end{center}

\begin{abstract}
We discuss a continuous variables method of quantum key distribution employing strongly polarized coherent 
states of light. The key encoding is performed using the variables known as Stokes parameters, 
rather than the field quadratures. Their quantum counterpart, the Stokes operators $\hat{S}_i$ (i=1,2,3), 
constitute a set of non-commuting operators, being the precision of simultaneous measurements of 
a pair of them limited by an uncertainty-like relation. Alice transmits a conveniently modulated two-mode 
coherent state, and Bob randomly measures one of the Stokes parameters of the incoming beam. After performing 
reconciliation and privacy amplification procedures, it is possible to distill a secret common key. We also
consider a non-ideal situation, in which coherent states with thermal noise, instead of pure coherent states, 
are used for encoding.
\end{abstract}

\section{Introduction}
\label{sec:Introduction}

Since the introduction of quantum cryptography by Bennett and Brassard more than twenty years ago \cite{bennett84}, 
research in this field has undergone a significant growth. As recent developments related to applications we may cite
the first bank transfer via a quantum protocol \cite{poppe04}, and the demonstration of quantum key 
distribution in telecom fibers over 122 km \cite{shields04}. Despite of the fact that quantum cryptography represents 
one of the most advanced sub-fields of quantum information processing, there is a great deal of research to be 
done. One of the main challenges in this respect is the integration of quantum 
cryptography protocols with more conventional communication systems, which would require systems robust enough to 
resist possible attacks as well as environmental influences. A considerable number of proposals, 
including the first one, the BB84 \cite{bennett84}, rely upon the existence of (low speed) single photon sources 
for the transmitter Alice and single photon detectors (with long recovery times) for the receiver Bob, being one of the 
forms of the protocol based on the polarization properties of single photons. Because the generation of single 
photons is not a simple task, most of the experiments use weak coherent pulses (highly attenuated lasers) instead,
as an approximation to single photons \cite{gisin02}. Some alternative proposals using continuous variables sources,
such as squeezed states have been already presented \cite{rhc00}. Although squeezed states-based schemes are 
interesting because they employ pulses with many photons instead of single 
photons\footnote{This could allow higher 
transmission rates, for instance.}, they still require special sources of non-classical light. Nevertheless, it has 
been shown that equivalent levels of security may be achieved using the ``quasi-classical'' coherent states 
\cite{grangier02,grangier03}, with full exploration of the continuous variable nature of coherent states (all 
continuous scheme). Coherent states are easy to generate, are relatively robust against losses, and therefore they
constitute a basic resource for viable quantum key distribution schemes. Usually, the key encoding in continuous variable schemes 
is implemented in the quadratures variables \cite{grangier02,grangier03}, 
being its security based on the impossibility of sharp simultaneous measurements of the quadratures themselves. 
However, a synchronized local oscillator for homodyne detection, connecting
Alice's and Bob's stations is required for quadrature measurements, which represents a difficulty if one aims 
a stable system for performing quantum key distribution between two points at larger distances. 
Here we discuss a different way of key encoding, yet using coherent states but without a 
necessity of a separate local oscillator. 
Our proposal \cite{borelli04} makes use of the polarization properties of coherent states. 
A polarized coherent state may be conveniently represented as a two-mode coherent state, or two single-mode 
coherent states excited in orthogonal directions \cite{silber02}. 
The variables which completely determine the polarization properties of the classical electromagnetic 
field are known as Stokes parameters \cite{stokes82,jackson99}, and their quantum mechanical analogues, 
the (hermitian) Stokes operators \cite{jauch59} are adequate tools for the quantum mechanical description of 
light polarization. Three of the four Stokes operators do not commute, yielding the well known \cite{silber02,jauch59}
uncertainty-like relations among them. 
As a matter of fact, any pair of non commuting quantum continuous variables (quadratures, 
polarization variables) would be suitable for a continuous variable quantum key distribution protocol. 
The expectation values as well as the variances of the Stokes operators may be readily 
measured using linear optical devices and PIN photodiodes, without the need of a separate local 
oscillator and single photon detectors \cite{silber02,agarwal03}. 
The quantum mechanical fluctuations of light polarization have been at the basis of 
proposals for continuous variable quantum cryptography using polarization-entangled beams \cite{silber02}, for instance. 
A different scheme using polarized laser beams and Stokes parameters measurements, without the need of entangled or squeezed states, 
has been recently implemented \cite{korolkova04}; it employs a predetermined set of four coherent states with distinct polarizations, 
and a post-selection procedure should introduced in order to reduce the information available to Eve. 
Another significant advance in continuous variable quantum key distribution is the recently 
proposed method which does not require random switching between measurement bases \cite{lam04}. 
As a result a significantly larger secret key rate is achieved, in comparison to the random switching case. 
Our paper is organized as follows; in section 2 we introduce the Stokes operators; 
in section 3 we show the basic steps to be followed in order to implement the protocol; in section 4 we discuss 
the security of the protocol against noise; in section 5 we summarize our conclusions.

\section{Stokes operators}
\label{sec:StokesOperators}

We consider two orthogonal modes of the electromagnetic field having polarizations 
oriented along the cartesian axis $x$ and $y$. In the case of the quantized field, the  
photon creation (annihilation) operators associated to each mode may be written as 
$\hat{a}^{\dagger}_{x}$ ($\hat{a}_{x}$) and $\hat{a}^{\dagger}_{y}$ ($\hat{a}_{y}$). They satisfy
the usual commutation relations

\begin{equation}
\left[\hat{a}_{i},\hat{a}^{\dagger}_{j}\right]=\delta_{ij},\;\;i,j=x,y.
\end{equation}

The hermitian Stokes operators are defined as \cite{jauch59}

\begin{equation} \hat{S}_{0}= \hat{a}^{\dagger}_{x}\hat{a}_{x} +
\hat{a}^{\dagger}_{y}\hat{a}_{y}=\hat{n}_{x}+\hat{n}_{y}
\end{equation}

\begin{equation} \hat{S}_{1}= \hat{a}^{\dagger}_{x}\hat{a}_{x} -
\hat{a}^{\dagger}_{y}\hat{a}_{y}=\hat{n}_{x}-\hat{n}_{y}
\end{equation}

\begin{equation} \hat{S}_{2}= \hat{a}^{\dagger}_{x}\hat{a}_{y} +
\hat{a}^{\dagger}_{y}\hat{a}_{x}
\end{equation}

\begin{equation} \hat{S}_{3}= i\left(\hat{a}^{\dagger}_{y}\hat{a}_{x} -
\hat{a}^{\dagger}_{x}\hat{a}_{y}\right).
\end{equation}

Consider the field prepared in the two-mode coherent state 

\begin{equation}
\left|\psi_{xy}\right>=\left|\alpha_{x}\right>_{x}\left|\alpha_{y}\right>_{y}=\hat{D}_{x}
\left(\alpha_{x}\right)\hat{D}_{y}
\left(\alpha_{y}\right)\left|0\right>_{x}\left|0\right>_{y},\label{twomodecoh}
\end{equation}
i.e., mode $x$ prepared in a single-mode coherent state $\left|\alpha_{x}\right>$ and mode 
$y$ prepared in a single-mode coherent state $\left|\alpha_{y}\right>$, with 
$\hat{a}_i\left|\alpha_i\right>=\alpha_i\left|\alpha_i\right>$. 
The expectation values of the Stokes operators of the field
in state $\left|\psi_{xy}\right>$ will read
\begin{eqnarray}
	\left<\hat{S}_{0}\right>&=&|\alpha_x|^2 + |\alpha_y|^2 \nonumber\\ 
	\left<\hat{S}_{1}\right>&=&|\alpha_x|^2 - |\alpha_y|^2 \nonumber\\ 
	\left<\hat{S}_{2}\right>&=& \alpha^*_x \alpha_y + \alpha^*_y \alpha_x\nonumber\\
	\left<\hat{S}_{3}\right>&=& i \left( \alpha^*_y \alpha_x - \alpha^*_x \alpha_y \right)
	\label{stokesexpec}
\end{eqnarray}
which are just the classical Stokes parameters describing the polarization of a light beam of intensity
$\left<\hat{S}_0\right>$ and with
an electric field with amplitude $\alpha_x$ in the $x$ direction and $\alpha_y$ in the $y$ direction. 
Therefore the coherent states correspond to classical fully polarized fields. 
However, the Stokes operators exhibit quantum mechanical fluctuations,
e.g., their variances in the two-mode coherent state above are

\begin{equation}
V_i \equiv \left<(\Delta \hat{S}_i)^2\right> = \left<\hat{S}_{i}^2\right> - \left<\hat{S}_{i}\right>^2 = 
\left<\hat{S}_{0}\right>. \ \ \ \ \ \ \ i=0,1,2,3
\end{equation}
The Stokes operators obey the following angular momentum-like commutation relations \cite{silber02,jauch59} 
\begin{equation}
\left[\hat{S}_{j},\hat{S}_{k}\right]=2i\epsilon_{jkl}\hat{S}_{l},\;\;j,k,l=1,2,3.
\end{equation}
This means that the precision of simultaneous measurements of a pair of Stokes parameters is limited by an uncertainty-like relation. For instance,  
\begin{equation}
\left(V_2 V_3\right)^{1/2} \geq \left|\left<\hat{S}_{1}\right>\right|.
\label{uncert}
\end{equation}
We remark that unlike the quadrature operators, the product of the variances of the Stokes operators is not a 
constant. However, for $\hat{S}_{0}$, whose mean value is basically the field intensity,
\begin{equation}
\left[\hat{S}_{0},\hat{S}_{i}\right]=0,\;\;i=1,2,3.
\end{equation}
The quantum mechanical properties of the Stokes operators turn them potential candidates for quantum key distribution
purposes. A most remarkable property of the Stokes operators is that their measurement may be accomplished with 
well established optical measurement methods, given the simple interpretation of the expectation values in Eqs.
(\ref{stokesexpec}) as differences of light intensity of components oriented along the reference axes ($xy$),
$\left<\hat{S}_{1}\right> = I_x - I_y$, along the axes rotated of $\pi/4$
$\left<\hat{S}_{2}\right> = I_{45^{\circ}} - I_{-45^{\circ}}$,
and the intensity difference between right circularly and left circularly polarized light components
$\left<\hat{S}_{3}\right> = I_{\sigma_+} - I_{\sigma_-}$.

\section{The protocol}
\label{sec:TheProtocol}

We present now the principles of our protocol: Alice generates a coherent beam 
of intensity $S_0$. A convenient preparation for that beam is a highly polarized two-mode coherent state.
Using the representation of Eq. (\ref{twomodecoh}), if the beam is strongly polarized in the $x$ direction, so
that $\left|\alpha_x\right|^2 \gg \left|\alpha_y\right|^2$, it follows from Eqs. (\ref{stokesexpec}) that
\begin{equation}
\left<\hat{S}_{1}\right> \approx \left<\hat{S}_{0}\right> = \left|\alpha_x\right|^2 . 
\end{equation}
We also have that $\left<\hat{S}_{1}\right> \gg \left<\hat{S}_{2}\right>, \left<\hat{S}_{3}\right>$. 
Therefore the uncertainty-like relation in Eq. (\ref{uncert}) may be approximately written as
\begin{equation}
\left(V_2 V_3\right)^{1/2} \geq |\alpha_x|^2.
\end{equation}
This means that for a given field intensity $|\alpha_x|^2$, the product of the variances
of $\hat{S}_2$ and $\hat{S}_3$ is basically a constant. 
As a consequence, we may also write
\begin{equation}
\left[\hat{S}_{2},\hat{S}_{3}\right]=2i|\alpha_x|^2.
\end{equation}
We recall that for the quadrature operators $\hat{X}=(\hat{a}^\dagger + \hat{a})$ and
$\hat{Y}=(\hat{a}^\dagger - \hat{a})i$, we have that $\left[\hat{X},\hat{Y}\right]=2i$. It would be then  
convenient to normalize the Stokes operators as $\hat{s}_i = \hat{S}_i/|\alpha_x|$, 
so that $\left[\hat{s}_{2},\hat{s}_{3}\right]=2i$. In summary, we have a clear correspondence between the noise
properties of the pair of non-commuting quadrature operators $(\hat{X},\hat{Y})$ and the pair of (normalized) 
Stokes operators $(\hat{s}_2,\hat{s}_3)$. Under such circumstances it becomes possible to encode the key 
elements in the Stokes variables ${S}_2$ and ${S}_3$ similarly as it is done in other continuous variables 
schemes based on the quadratures, for instance \cite{grangier02}. 
The beam generated by Alice crosses an electro-optical modulator
and a magneto-optical modulator in sequence, in such a way that small random and independent modulations 
of the Stokes variables $S_3$ and $S_2$ are performed, i.e., two random numbers, $S_3$ and $S_2$, are 
drawn from a Gaussian distribution of mean value zero and variance
$V_m$ \cite{grangier02,grangier03}. Alice then sends the modulated beam to
Bob, who randomly chooses to measure either $S_2$ or $S_3$. As it is usual in quantum cryptography 
protocols, Alice and Bob establish communication via a public authenticated channel and Bob informs
Alice which Stokes variable he has measured. After repeating that process several times, Alice and 
Bob will share a set of {\it Gaussian correlated variables}, or ``{\it key elements}'' 
\cite{grangier03}. Such raw data must be adequately processed in order to generate a common 
secret binary key (a string of bits). Error correction should be performed, and the information available to
a potential eavesdropper (Eve) should be minimized via a ``sliced reconciliation procedure'' \cite{cerf04}, 
necessary to convert the continuous correlated variables in bit strings; after that the bit string should
be made secret (privacy amplification). 

As we have already seen, the noise properties of the Stokes variables $S_2$ and $S_3$ of
a strongly linearly polarized coherent beam are similar to the ones of the quadrature
variables. We therefore expect that the procedures and analysis of the protocol security to be 
similar as well. A key point concerning
the security of our protocol is the random Gaussian modulation of the Stokes parameters. The Gaussian
noise superimposed to the quantum noise of the polarization variables makes virtually impossible the
retrieval of any information by Eve, while the reconciliation procedure allows to both Alice and Bob
to establish a common secret key. This means that the data processing necessary for the key 
establishment is basically the same to the one implemented in the quadrature protocol \cite{grangier03}.

\section{Security and noisy states}
\label{sec:Security}

The eavesdropper Eve is supposed to have perfect equipment and unlimited computational power at her disposal. 
This means that she may perform different kinds of attacks while tapping the quantum channel between Alice and Bob.
Here we are going to analize a particular (but rather effective) attack known as the individual cloning attack.
We assume that Eve is able to intercept the signal producing two identical copies of the quantum state sent by
Alice. She keeps one copy and sends the other to Bob. However, because of the no-cloning theorem
\cite{wootters82,cerf00}, i.e., the impossibility of making perfect copies from unknown states, 
neither Eve nor Bob will
be able to retrieve all the information transmitted by Alice. In other words, the cloning process will introduce 
noise in the relevant variables. Moreover, the measurement of one variable, e.g., $s_2^E$ by Eve, will affect the
values of $s_3^B$ measured by Bob. If Alice prepares and sends a beam having a Stokes parameter $\hat{s}_2^A$ 
$(\hat{s}_3^A)$, Eve will keep a copy with Stokes parameter $\hat{s}_2^E$ $(\hat{s}_3^E)$ with channel
noise $\hat{E}_{s_2}(\hat{E}_{s_3})$, and Bob will receive one copy with a Stokes parameter $\hat{s}_2^B$ 
$(\hat{s}_3^B)$ with channel noise $\hat{B}_{s_2}(\hat{B}_{s_3})$. 
Taking into account the commutation relations between the (normalized) Stokes parameters, and assuming that the noises 
are not correlated to the input signals, it is straightforward \cite{grangier01} to obtain the following crossed 
uncertainty relation relative to the noise variances\footnote{It is convenient to work with 
the variances normalized to $|\alpha_x|$, or $v_i = V_i/|\alpha_x|^2$. }
\begin{equation}
\left(v_{B;s_2} v_{E;s_3}\right)^{1/2} \geq 1,\label{crossed1}
\end{equation}
and
\begin{equation}
\left(v_{B;s_3} v_{E;s_2}\right)^{1/2} \geq 1.
\end{equation}
This means that, because of the constraints above, even a little noise $v_{E;s_3}$($v_{E;s_2}$)
on Eve's copy will cause a large disturbance $v_{B;s_2}$($v_{B;s_3}$) on Bob's one, if they
measure different Stokes variables ($s_3$ and $s_2$, respectively). 
In the individual cloning attack, Eve keeps her copy until the reconciliation process, i.e., when the public 
announcement of measurements is made, and only then measuring the ``correct'' Stokes variable as an attempt to 
retrieve useful information. 
The Shannon information theory is particularly suitable if one deals with Gaussian continuous 
variables such as coherent states transmitted through a Gaussian noisy channel; an important 
property is that Gaussian noises from independent sources are {\it additive}. Moreover, a Gaussian 
modulation of the signal sent by Alice allows the reduction of the information available to
Eve \cite{grangier02,grangier03}. In this formalism, the optimum mutual information between Alice and Bob $I_{AB}$
is given by Shannon's formula for a noisy (Gaussian) transmission channel \cite{shannon48}
\begin{equation}
I_{AB} = \frac{1}{2}\log_2\left(1+\frac{v_m}{v^B_n}\right) 
\end{equation}
where $v_m$ is the (normalized) variance of the signal and $v^B_n$ is the noise variance.
If one wants to perform reliable quantum key distribution in the presence of Eve, the mutual information 
between Alice and Bob $I_{AB}$ must be larger than the mutual information between Alice and Eve $I_{AE}$
\cite{korner78}, i.e.,
\begin{equation}
\Delta I = I_{AB} - I_{AE} > 0.
\end{equation}
The Stokes parameters are modulated by Alice with a variance $v_m$.  
As soon as the beam enters a Gaussian noisy channel, the signal received by Bob will acquire noise with variance 
$v_{B;s_i}$ in its $i$-th Stokes parameter. We can make, for simplicity, $v_{B;s_2} = v_{B;s_3} = n_B $. 
We also have to take into account the intrinsic quantum noise in the Stokes parameters, so that $v^B_n = 1 + n_B$. Therefore 
\begin{equation}
I_{AB} = \frac{1}{2}\log_2\left(1+\frac{v_m}{v^B_n}\right) = 
\frac{1}{2}\log_2\left(\frac{v_m + 1 + n_B}{1 + n_B}\right).
\end{equation}
Similarly, for $I_{AE}$, $v^E_n = 1 + n_E$. Note that, because of Eq.(\ref{crossed1}), $n_E = 1/n_B$, and thus
\begin{equation}
I_{AE} = \frac{1}{2}\log_2\left(1+\frac{v_m}{v^E_n}\right) = 
\frac{1}{2}\log_2\left(\frac{v_m + 1 + 1/n_B}{1 + 1/n_B}\right).
\end{equation}
We have that as long as $n_B < 1$, $\Delta I = I_{AB} - I_{AE}$ increases with the modulation variance $v_m$,
analogously to the quadrature protocol treated in references 
\cite{grangier02,grangier03}, but for the Stokes variables, instead. The line transmission parameter 
is given by $\eta = 1/(1+n_B)$ \cite{grangier02}; under the condition $n_B < 1$, secure key 
distribution is allowed in a transmission line with losses less than 50\% \cite{grangier02}. 
Such a result stands for ``direct reconciliation'' protocols, where Alice sends to Bob information
via a classical public channel. The 50\% limit (3 dB) may be overcome using a ``reverse reconciliation technique'' 
\cite{grangier03}. All noise in the transmission process, apart 
from the handy Gaussian modulation of the signal and the state intrinsic quantum noise is attributed to Eve. 
The preceding discussion has been based on the transmission of fields prepared in pure coherent states. 
Nevertheless, a quantum field is more generally (and realistically) described as a statistical mixture, specially if there are imperfections during the state preparation \cite{imoto}. Now we are going to discuss the case in which noisy
mixed states, rather than pure coherent states are transmitted. We consider that the states to be transmitted through 
the quantum channel are a distribution (statistical mixture) of coherent states with Gaussian weight (thermal 
coherent states), having density operator
\begin{equation}
\hat\rho_{tc}=\int_{-\infty}^{\infty}\int_{-\infty}^{\infty}d^{2}\!\alpha_{1}d^{2}\!\alpha_{2}
\ P(\alpha_{1}\alpha_{2})\
|\alpha_{1}\alpha_{2}\rangle\langle\alpha_{1}\alpha_{2}|,\label{termocoe}
\end{equation}
with
\begin{equation}
P(\alpha_{1};\alpha_{2})=\frac{1}{\pi^{2}\,\overline{n}_{th}^2}
\exp\left({-\frac{|\alpha_{1}-\alpha_{0x}|^2}{\overline{n}_{th}}-\frac{|\alpha_{2}-\alpha_{0y}|^2}{\overline{n}_{th}}}
\right),
\end{equation}
and where $\overline{n}_{th}$ is the mean number of thermal photons in each mode (the same for simplicity) 
and $\alpha_{0i}$ is the central coherent amplitude in the $i$-th mode of the field. In the case of a
strongly polarized beam, or $|\alpha_{0x}|^2 \gg |\alpha_{0y}|^2$, and if we neglect the thermal
fluctuations in the $y$ mode, the variances of the Stokes parameters $\hat{S}_2$ and $\hat{S}_3$ 
will be
\begin{equation}
V_{2}=V_{3}=\overline{n}_{th}+|\alpha_{0x}|^2.
\end{equation}
We may again employ the normalized variances $v_i = V_i/|\alpha_{0x}|^2$, so that
\begin{equation}
v_{2} = v_{3}= 1 + r,
\end{equation}
where $r = \overline{n}_{th}/|\alpha_{0x}|^2$ is a dimensionless parameter indicating the amount of  
thermal noise. Bearing in mind that the transmitted field is a quantum Gaussian state, the noises are additive, and 
we may write the corresponding mutual informations as
\begin{equation}
I_{AB}=\frac{1}{2}\log\left[\frac{v_{m} + (1+r) + n_B}{(1+r) + n_B}\right],
\end{equation}
\begin{equation}
I_{AE}=\frac{1}{2}\log\left[\frac{v_{m} + (1+r) + 1/n_B}{(1+r) + 1/n_B}\right].
\end{equation}
The influence of thermal noise is graphically shown in figure 1, where it is shown a plot of $\Delta I$
as a function of the noise parameter $r$ and the line transmission $\eta$. As one would expect, the secret
bit rate $\Delta I$ decreases for noisier fields (larger $r$), reducing the secure transmission range of the protocol. 
Nevertheless, even for relatively noisy fields, it is still possible secure transmission in a line having up to 50\% of losses
(using direct reconciliation), as in the case of pure coherent fields ($r=0$). 
We would like to remark that the consequences of the thermal noise 
considered here, not accessible to Eve (not general), are normally overlooked in the existing literature, and
the inclusion of such a noise as we did in our analysis is surely of relevance if one thinks about a more realistic physical situation. 

\begin{figure}[ht]
\begin{center}
\includegraphics{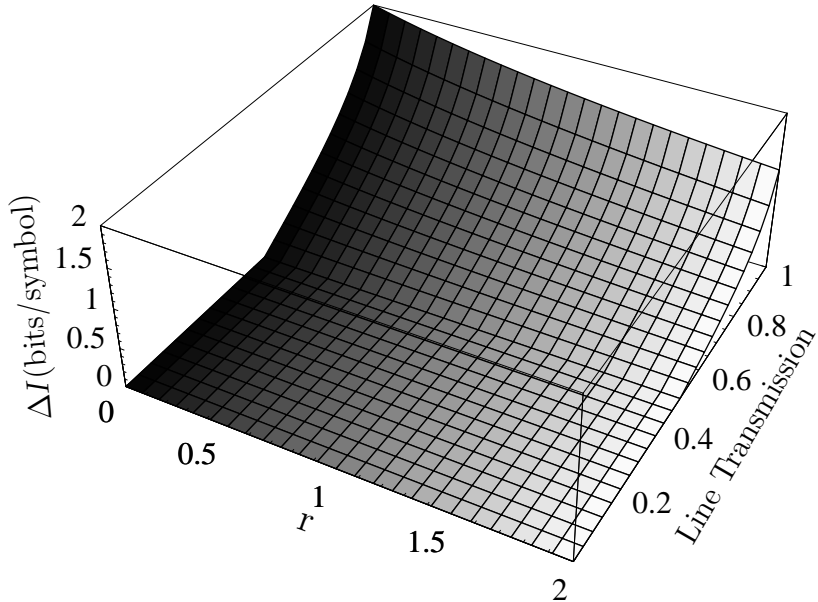}
\end{center}
\caption{\label{figure1} Secret key bit rate as a function of the noise parameter $r$ and the line transmission
$\eta$.}
\end{figure}
\section{Conclusion}
\label{sec:Conclusion}

The continuous variables quantum key distribution protocols belong to an emerging field; there is great potential
for further development of these methods towards applications. There is no need of special sources 
for single photons or entangled beams, and in the case of polarization encoding, precise 
synchronization of distant stations is not necessary, in contrast to other protocols. 
The protocol presented here is an {\it all continuous variables} one, in the spirit of the quadrature operators 
method \cite{grangier03}, but without the need of synchronized stations for homodyne detection. 
A quantum cryptography protocol based on polarized coherent states as presented here might be suitable for 
integration to other cryptographic systems, e.g., it could be used to provide secret seed keys to a 
cryptography protocol \cite{barbosa03} using two mode coherent states. 
In conclusion, we have presented an alternative method for continuous variable quantum key distribution which makes 
use of the polarization properties of coherent states. The encoding in the polarization variables of coherent states 
offers several advantages; coherent states are easily generated, while the Stokes parameters are easily measured. 
Further studies on the protocol as well as its experimental implementation are being carried out and 
will be considered elsewhere.

\section*{Acknowledgements}

The authors would like to thank L.S. Aguiar for useful discussions. This work is partially supported by CNPq 
(Conselho Nacional para o Desenvolvimento Cient\'\i fico e Tecnol\'ogico), CAPES (Coordena\c c\~ao de Aperfei\c coamento de Pessoal de N\'\i vel Superior) and FAPESP (Funda\c c\~ao 
de Amparo \`a Pesquisa do Estado de S\~ao Paulo), Brazil.

\end{document}